\documentclass[twocolumn,10pt]{article}
\usepackage{graphicx}

\begin{document}
\title{
 Violation of Cluster Property in Superconducting Qubit}

\author{Tomo Munehisa \\
Faculty of Engineering, University of Yamanashi,\\ Kofu, Japan, 400-8511,\\
 munehisa8tomo@gmail.com}

\maketitle

\begin{abstract}
Spontaneous symmetry breaking is well known to be 
a macroscopic phenomenon
 in quantum physics in  many body systems.
 The  essential  features of this phenomenon   are the energy degeneracy and 
 the existence of the local operator that can modify  the degenerate states.
Due to these properties 
 the lowest energy state is not  stable.  
In order to obtain the stable ground state
 we introduce the explicitly breaking interaction, by which it is difficult for 
 the local operator to modify the ground state.
 
In  the superconducting  qubit
there exists the stable eigenstate with the definite charge number.
We point out that in this system, the local operator can change  this state to another state.
As  results the cluster property violates, which means that
the events in the far distance are not irrelevant.
In this paper we  examine the cluster property 
in  the superconducting qubit.
Then we  propose a method by
 cavity quantum electrodynamics with two cavities in order to
observe  the violation.

\end{abstract}

\section{Introduction}
Spontaneous symmetry breaking (SSB) is a fundamental  concept in
 quantum physics, which
includes the condensed matter physics and particle physics.
The original discovery of this concept, superconductivity\cite{SSB1}, has widespread applications in social systems.
 While in particle physics, we give the mass to the elementary particles
 by SSB\cite{SSB2}.
We  state that SSB occurs if the ground state is not the eigenstate
of the symmetry charge.
This simple statement on SSB is ambiguous, because the energy eigenstate
is the eigenstate of charge in the finite size system.
Therefore, SSB occurs in the exact meanings, only when the system size is infinitely large.
However, even if the system has the finite size, we find the signals for SSB,
so that we make discussions on the system, whose size is quite large, but finite. 

To explain SSB, first we clarify the ground state or the vacuum state.
Its naive concept for it is 	`nothing'.
This concept can be replaced by the uniformity of the spatial translation.
It implies that you see the same scene wherever you go.
Therefore, when  the state is  uniform under the spatial translation,  there are  no particles or no localized excitations.
When there is   only one state that has the uniformity, i.e. we have a single ground state, this state is the eigenstate of  the charge. 
In this case we cannot find any peculiar phenomenon.

In the other hand, in SSB we find  many  states with the uniformity.
Then  energies of these states become degenerate when the system size  is infinitely large.
This degeneracy is the first feature of SSB.
Another ingredients of SSB is the existence of 
   the local operator that can  change  the  state with the uniformity to another\cite{My1}\cite{My2}.
This existence implies that the vacuum state changes by the local operation.
In other word we can find correlations between the local operator and another
even if the distance  is quite large,
because at location $A$ the local operator changes
the state, and  another  operator changes it to the original state again at location $B$.
As results  the existence of the local operator
threatens the principle of the cluster property or the decomposition\cite{CP1}\cite{CP2}\cite{CP3}\cite{CP4}\cite{CP5}\cite{CP6}.

 The next question is whether the lowest energy state is stable or unstable
 in the finite system.
 Answer is that this state is unstable.
 The reason is the following.
Even if  system size $N$ is finite, the lowest energy state is
unstable because the energy gap is of order of $N^{-1}$ and the local operators change the state.
Therefore, the lowest energy eigenstate changes to another when we include
small local interactions.

It is well known that the lowest energy state becomes stable by including the explicitly  breaking interaction into Hamiltonian\cite{My1}\cite{My2}\cite{CPx1}\cite{CPx2}. %
By this interaction the lowest energy state 
 has no definite quantum number.
  Then  the violation of the cluster property (VCP) 
  becomes tinny effects\cite{My1}\cite{My2}.
 Note that VCP can be observed in the indirect way\cite{My3}.

When we include  operator $\hat{Q}^2$ of the charge squared  into Hamiltonian, the degeneracy disappears and the lowest energy
state with a definite  charge number becomes stable even if there exists a small interaction breaking the symmetry.
Since this operator $\hat{Q}^2$  keeps the ability of
 the local operator to change the lowest energy state into the excited  state, we can observe VCP.

In recent years  the experimental researchers\cite{qbit0} have produced the superconducting  system that contains  operator $\hat{Q}^2$ of the charge squared so that
  the lowest energy state has the definite charge number $n_Q$.
This state and the first excited state with  $n_Q+1$ 
 are formed as  a quantum bit (qubit)\cite{qbit5}\cite{qbit6}\cite{qbit7}.
 Specially this is  called as the charge qubit since the state has
 the definite charge number.
 There is active research underway to develop quantum computing using charge qubits\cite{qbitReview1}\cite{qbitReview2}\cite{qbitReview3}\cite{qbitReview4}\cite{QC1}.

In this paper
we  study  the violation of cluster property in
 the  superconducting charge qubit.
Since the lowest energy state has the definite charge number $n_Q$,
we find that the  cluster property violates at  finite magnitude, 
and it does not depend on the size of the system.
We apply cavity quantum electrodynamics(cQED)\cite{cQED1}
to observe  the violation.
In  cQED
the electromagnetic field confined in a cavity couples with matter as atoms. 
By the  coupling with the matter,  the resonance frequency of the photon changes.
Since  the  measurement on the frequency is quite execute,
we find the effect even if  the coupling is quite small\cite{cQED1}.

 In the superconducting qubit,
we find quite active studies on cQED
for  detecting the state\cite{cQED2} or for making high coherence in Josephson junction qubits\cite{cQED3}.
For our purpose to observe the violation, we propose a method by cQED with two cavities in the superconducting qubit.
VCP means that the correlation of two operators is finite, even if
they are separated at the far distance.
Therefore,  we study  effects by 
the  correlation of photon operators in two cavities, one of which  is far from another.

Contents of this paper are as follows. In the next section we present a review
on the states in SSB and VCP. Here we assume U(1) symmetry, whose eigen value is one integer $n_Q$. Then we show VCP in the superconducting qubit.
In section 3 we  introduce  the extended Jaynes-Cummings model\cite{JC1}\cite{JC2},  in order to study  two photons at  cavities in  cQED in the qubit states.
Since the coupling of the photon with the superconducting qubit
is small, we examine two cases.
In the first case  the photon energy equals to the energy gap in the qubit.
In another case there is no equality, where we present the effective
Hamiltonian.
The final section is devoted to summary and discussion. 

In appendix A we show a method to observe VCP indirectly.
In a next appendix  we present eigenvalues and eigenstates when
energies of two photons agree with the gap energy in the superconducting. 
 In appendix C we will derive the effective
Hamiltonian for the small coupling of the photon with
the superconducting. Also, we show the eigenvalues for the effective
Hamiltonian and examine the time evolution of the photon number operators.
\vskip 1cm


\section{Violation of Cluster Property}
In field theory or many-body problems, a   local operator $\hat{\phi}(\vec{x})$ is a basic element  in the theory, where  $\vec{x}$ denotes a position in space.
Also, we suppose that Hamiltonian $\hat{H}$  does not depend on
the specific location, which means that  Hamiltonian $\hat{H}$ is invariant under  space transformation $\hat{T}(\vec{a})$.
This is defined by
\begin{eqnarray}
\hat{T}(\vec{a}) \hat{\phi}(\vec{x}) \hat{T}^\dagger(\vec{a})\equiv\hat{\phi}(\vec{x}+\vec{a} ) \ \ , \ \ 
[\hat{T}(\vec{a}), \hat{H}]=0
 \ \ .\label{Vc0a}
\end{eqnarray}
Here we assume that the ground state $ |G\rangle $ 
is uniform under the translation, that is $\hat{T}(\vec{a}) |G\rangle = |G\rangle $. 
We consider that the operator $\hat{\phi}(\vec{x})$ has the continuous symmetry,
and we  introduce  charge operator $\hat{Q}$ which is defined by
\begin{eqnarray}
\exp(i\alpha \hat{Q}) \hat{\phi}(\vec{x}) \exp(-i\alpha \hat{Q})
\equiv \hat{\phi}(\vec{x}) e^{-i\alpha}
 \ \ .\label{Vc0b}
\end{eqnarray}
Here we consider $U(1)$ symmetry for concrete descriptions hereafter.
Then we assume that Hamiltonian $\hat{H}$ is invariant under this transformation,  $ [\hat{H},\hat{Q}]=0$.
Therefore, the energy eigen state has the quantum number $n_Q$ of  $\hat{Q}$.

The continuous symmetry spontaneously breaks,
or spontaneously symmetry  breaking (SSB) occurs,
if  $ |G\rangle $ does not have a definite number of $n_Q$,
which is expressed by
\begin{eqnarray}
\hat{Q} |G\rangle\not= |G\rangle n_Q
 \ \ .\label{Vc0c}
\end{eqnarray}
More precisely the ground state $ |G\rangle $ has the non-zero expectation value $v\equiv \langle G|\hat{\phi}^\dagger(\vec{x})| G\rangle \not=0 $.

In the above statements there is some ambiguity, because the lowest energy state has the definite number $n_Q$,   when  system size $N$ is 
 finite. Therefore, we cannot find the ground state given in  (\ref{Vc0c})
 for  finite $N$. 
In order to discuss SSB definitely, we suppose that
$N$ is quite large, but is not infinitely large.
Here  by system size $N$  we define a number of independent local operators $\hat{\phi}(\vec{x})$.
Also, we use the word `the lowest energy state' instead of `ground state'
for clear discussions. 

For finite $N$,  we make $\hat{Q}$  well defined, by which
the energy eigenstate has a definite number $n_Q$ for $\hat{Q}$.
It implies that
\begin{eqnarray}
\hat{H}| E,n_Q\rangle= | E,n_Q\rangle E \ \  , \nonumber \\
\hat{Q}| E,n_Q\rangle= | E,n_Q\rangle n_Q
 \ \ .\label{Vc2a}
\end{eqnarray}
Since we define  the translation symmetry in  finite $N$ system,  we assume that the lowest energy state
is uniform. Here we discuss the uniform state only, which is denoted by  $|n_Q\rangle$.
\begin{eqnarray}
\hat{H}| n_Q\rangle= | n_Q\rangle  E_{n_Q} \ \ , \ \ 
\hat{T}| n_Q\rangle= | n_Q\rangle
 \ \ .\label{Vc2b}
\end{eqnarray}
SSB  will occur when the following two properties hold.
\begin{eqnarray}
\Delta_{n_Q}\equiv  E_{n_Q+1} -E_{n_Q}\ \ , \ \  \lim_{N\rightarrow \infty}\Delta_{n_Q}=0
 \ \  , \label{Vc2c} \\
 v_{n_Q}=\langle n_Q+1|\hat{\phi}^\dagger(\vec{x})| n_Q\rangle
 \ \ ,  \ \  \lim_{N\rightarrow \infty}v_{n_Q}=v\not=0
 \ \ .\label{Vc2d}
\end{eqnarray}
For anti-ferromagnets and the non-liner  sigma models we find that $\Delta_{n_Q}=an_Q^2/N$ with some constant $a$.
For SSB we need clearly the first property (\ref{Vc2c}). Because
when there exists no degeneracy,
 we have the state $ | n_Q\rangle$ with the lowest $ E_{n_Q}$,
 and this has the eigen value $n_Q$.
 
 The second property (\ref{Vc2d}) is required for SSB.
 If $v=0$, the lowest energy eigen state with $n_Q$ is stable.
 Therefore, by two properties (\ref{Vc2c}), (\ref{Vc2d}),
 $| n_Q\rangle$'s are unstable so that
 the lowest energy state is a coherent state $ \sum_{n_Q}| n_Q\rangle c_{n_Q}$
 when there exists a symmetry breaking interaction.
 Note the  coherent state has the property (\ref{Vc0c}).
 This stability will be discussed after we argue the cluster property.

Here we will point that the cluster property violates
by the second condition(\ref{Vc2d}).
In order to understand the violation,
first we note that the local operator $\hat{\phi}(\vec{x})$ can change the eigen state
  $| n_Q\rangle$ to $| n_Q-1\rangle$.
Therefore, we find the following.
\begin{eqnarray}
\langle n_Q|\hat{\phi}(\vec{x}) \hat{\phi}^\dagger(\vec{y})| n_Q\rangle\nonumber\\
=\langle n_Q|\hat{\phi}(\vec{x}) | n_Q+1\rangle \langle n_Q+1|   \hat{\phi}^\dagger(\vec{y})| n_Q\rangle
 \nonumber\\
+{\rm other\ contribution}\nonumber\\
 =v_{n_Q}^*v_{n_Q}+{\rm other\ contribution}
 \ \ .\label{Vc1a}
\end{eqnarray}
Here  other contribution decreases when $\vec{x}$ goes away from $\vec{y}$.
Even when $\vec{x} $ is quite far from $\vec{y}$, the following product does not vanish by (\ref{Vc2a}).
\begin{eqnarray}
\lim_{| \vec{x}-\vec{y}  |:{\rm large}}\langle n_Q|\hat{\phi}(\vec{x}) \hat{\phi}^\dagger(\vec{y})| n_Q\rangle=|v_{n_Q}|^2
 \ \ .\label{Vc1c}
\end{eqnarray}
Since $\langle n_Q|\hat{\phi}(\vec{x})| n_Q\rangle=0$,
we have
\begin{eqnarray}
\lim_{| \vec{x}-\vec{y}  |:{\rm large}}\{\langle n_Q|(\hat{\phi}(\vec{x})-\langle n_Q|\hat{\phi}(\vec{x})| n_Q\rangle)  \nonumber \\
(\hat{\phi}^\dagger(\vec{y})-\langle n_Q|\hat{\phi}^\dagger(\vec{y})| n_Q\rangle)
 | n_Q\rangle\}=v_{n_Q}^2
 \ \ .\label{Vc1b}
\end{eqnarray}
When the cluster property  holds for $|n_Q\rangle$, 
the value of the right hand-side must vanish.
Therefore, we conclude that the cluster property violates for the eigen state
$| n_Q\rangle$ by the property (\ref{Vc2d}).

Is the lowest energy state $| n_Q\rangle$ stable  in the real material?
Here we will make clear the concept of stability.
Theorists suggest Hamiltonian in order to describe the experimental fact.
This Hamiltonian is quite simple because we have to solve the eigenvalue
equation for the Hamiltonian.
When we say that the material has a symmetry,
we assume to neglect effects by explicitly symmetry breaking
 interactions. Then we construct Hamiltonian that has
 symmetry.
The proposed Hamiltonian will only be valid after verifying that the results are consistent.
 One requirement for this consistency 
 is the stability of the lowest energy state.
Here we argue that the lowest energy state is stable, only when
this state  changes little  even when  small interactions are imposed.
We will show more  precise statement.
When we include  the interaction whose  strength is $f$,
we find a new state $| n_Q\rangle_{new}=c_0| n_Q\rangle
+c_{other}| {\rm other}\rangle$, where$ | {\rm other}\rangle$
differs from $| n_Q\rangle$,
We define that    the lowest energy state is stable, when    $c_{other}$ is of order of $f$.
By this definition of the stability, the lowest energy state is stable, when
there exists a finite energy gap.

The above  definition means that the state $|n_Q\rangle$ is not stable.
Let us consider the following interaction.
\begin{eqnarray}
\hat{H}_b=f\sum_{\vec{x}}\{ \hat{\phi}(\vec{x})+\hat{\phi}^\dagger(\vec{x})\}
 \ \ .\label{Hb1}
\end{eqnarray}
Then we find that
the lowest energy state becomes $\sum_{n_Q} |n_Q\rangle c_{n_Q}$, where
$c_{n_Q}$ is irrelevant with $f$, if $f$ is larger than the energy gap
$\propto 1/N$.

In our work\cite{My1} on the quantum spin models, 
we examined the stability of the lowest energy state when we include the interaction (\ref{Hb1}) to original Hamiltonian which has the continuous symmetry.
In this Hamiltonian
 the energy gap is of order of $\sqrt{f}$, by which the lowest energy state
 is stable even if  there exists other interaction with its strength smaller than
 $f$.
The local operator  $\hat{\phi}(\vec{x})$ can change
the lowest energy  state to the second state, but this magnitude is 
 $1/\sqrt{f^{1/2}N}$.
By these estimates we conclude that the lowest energy state  
$\sum_{n_Q} |n_Q\rangle c_{n_Q}$ is stable
even for the small $f$ when  size $N$ is quite large.
By the stability we find that 
the violation of the cluster property is order of $1/(N\sqrt{f})$
so that the violation is a tinny effect.

Here we point out that there is  the indirect method to observe the effect by the violation\cite{My3}.
We will consider an interaction  $\hat{H}_{ob}$
in order to observe VCP.
\begin{eqnarray}
\hat{H}_{ ob}=g_{ ob}\{\hat{O}(\vec{x}) \hat{\phi}(\vec{x})+ 
\hat{O}(\vec{y}) \hat{\phi}(\vec{y})
+c.c.\}
 \ \ .\label{Vc4b}
\end{eqnarray}
Here the operator $\hat{O}(\vec{x}) $ is consisted of other degree of freedom which
differs from $\hat{\phi}(\vec{x})$.
The correlation between $\hat{O}(\vec{x}) $ and $\hat{O}(\vec{y}) $ 
is the same as that between $\hat{\phi}(\vec{x}) $ and $\hat{\phi}(\vec{y}) $,
so that this correlation 
is independent of the distance $ |\vec{x}-\vec{y}|$.
%
The detailed discussion will be found in appendix A.

In superconducting circuits\cite{qbit0}\cite{qbit5}\cite{qbit6}, it is now possible to include $(\hat{Q}-n_G)^2$ in the system's Hamiltonian.
This square operator resolves the difficulty of  the instability due to  degeneracy so that the eigenstate
 $| n_Q\rangle$ with the fixed number of the charge
 becomes  stable. As results  the lowest energy  state $| E_{lst}\rangle$
 has  the  definite number of the charge.
 
We will present  Hamiltonian of $(\hat{Q}-\bar{n}_G)^2$  constructed  by $| n_Q\rangle$,
 which is given by
\begin{eqnarray}
\hat{H}_{S}\equiv  b(\hat{Q}-\bar{n}_G)^2\sim  b\sum_{n_Q}(n_Q-\bar{n}_G)^2|n_Q\rangle \langle n_Q|
 \ \ .\label{Vc3a}
\end{eqnarray}
Here a parameter $b$ is independent on the size $N$.
For integer $n_Q^*$ that gives us  the minimum value of $(n_Q-\bar{n}_G)^2$, we have 
the lowest energy state 
$| E_{lst}\rangle =| n_Q^*\rangle $.
Then the local  operator  $\hat{\phi}^\dagger(\vec{x})$
can change the ground state into the excited state $| n_Q^*+1\rangle $.
Therefore, the magnitude of the violation of  the cluster property is not $O(1/N)$, but  $O(b)$.

The research target of the active studies on the superconducting circuits
is quantum computing, for which  we need the two levels states.
In order to obtain these states, we choose a parameter $\bar{n}_G=n_Q^*+1/2-\epsilon$, where $\epsilon$ is a small number.
In this case Hamiltonian becomes
\begin{eqnarray}
\hat{H}_{S_\epsilon}= b( \frac{1}{4}- \epsilon+\epsilon^2     ) | n_Q^*\rangle\langle n_Q^*|+\nonumber \\
 +b(\frac{1}{4}+ \epsilon+\epsilon^2     )| n_Q^*+1 \rangle \langle n_Q^*+1|+
\nonumber \\
+\sum_{n_Q\not=
n_Q^*,n_Q^*+1  }(n_Q-n_Q^*-\frac{1}{2}+ \epsilon)^2| n_Q\rangle \langle n_Q|
 \ \ .\label{Vc3b}
\end{eqnarray}
Clearly  the first excited  energy $E_{fe}=b( \frac{1}{4}+ \epsilon+\epsilon^2) $  is close to
 the lowest  energy $E_{lst}=b(\frac{1}{4}- \epsilon+\epsilon^2) $.
 The difference between them is $2b\epsilon$, while
 the difference between   the second excited  energy $E_{se} $ and $E_{lst}$ is $b(2-3\epsilon)$, which 
  is larger than  $2b\epsilon$, when $\epsilon$ is  small.
Therefore, we consider only two states $ |n_Q^*\rangle$ and $| n_Q^*+1\rangle$
effectively, which work as the two logic states(qubit)\cite{qbitReview2}.
\begin{eqnarray}
\hat{H}_{S_\epsilon}\longrightarrow \hat{H}_{qb}= b( \frac{1}{4}- \epsilon+\epsilon^2     ) | n_Q^*\rangle\langle n_Q^*|+\nonumber \\
 b( \frac{1}{4}+ \epsilon+\epsilon^2     )| n_Q^*+1 \rangle \langle n_Q^*+1|
 \ \ .\label{Vc3c}
\end{eqnarray}
Here we  call  these  states superconducting qubits.
On  operations of  the qubit state we can introduce Pauli matrices $\sigma^z,\sigma^x$. Using  the matrices we have 
\begin{eqnarray}
\hat{H}_{qb}=b\epsilon\sigma^z +b( \frac{1}{4}+\epsilon^2     ) 
 \ \ .\label{Vc3d}
\end{eqnarray}
Here the down state is $| n_Q^*\rangle$, while the up state is  
$| n_Q^*+1\rangle$.

Cavity quantum electrodynamics (cQED)\cite{cQED1} is the study of the interaction between light confined in a reflective cavity and atoms or other particles,
 under conditions where the quantum nature of photons is significant.
 By cQED we  probe the  state without disturbing the superconducting qubit\cite{cQED2}\cite{cQED3}.
 The cavity is  connected with the superconducting, and the photon in the cavity couples with the electric charge.
 If we denote $\hat{a}_{JC}$ as the annihilation operator of the photon, the interaction is given by Jaynes-Cummings model\cite{JC1}\cite{JC2} for the down state 
 $|n_Q^*\rangle$ and the up state $|n_Q^*+1\rangle$. The Hamiltonian of
 this model is given by
 \begin{eqnarray}
 \hat{H}_{JC}=\frac{\Omega}{2}\sigma^z + \omega  \hat{a}_{JC}^\dagger \hat{a}_{JC} + \nonumber \\
 g_{JC}(\hat{a}_{JC}\sigma^+ + \hat{a}_{JC}^\dagger\sigma^-)
  \ \ .\label{Vc4c}
\end{eqnarray}
 Although the coupling $g_{JC}$ is small, the interaction induces
 the fine change of the frequency of the photon, which can be measured  precisely.

 In order to study the cluster property,
we will apply cQED to the interaction (\ref{Vc4b}) so that   we use the photon operator $\hat{a}_{JC}(\vec{x})$   for $\hat{O}(\vec{x})$ and $\hat{a}_{JC}(\vec{y})$   for $\hat{O}(\vec{y})$.
 Here the coordinate $\vec{x}$ denotes the position of cavity $A$, 
  while the coordinate $\vec{y}$ denotes the position of cavity $B$.
  The position is  far from another. Therefore, we have the following interaction.
 \begin{eqnarray}
\hat{V}_{cQED}=g_p\{\hat{a}_{JC}^\dagger(\vec{x}) \hat{\phi}(\vec{x})+ 
\hat{a}_{JC}^\dagger(\vec{y}) \hat{\phi}(\vec{y})+c.c.\} \nonumber \\
=g_p\{\hat{a}^\dagger \hat{\phi}(\vec{x})+ 
\hat{b}^\dagger \hat{\phi}(\vec{y})+c.c.\}
 \ \ .\label{Vc4d}
\end{eqnarray}
Here we use $\hat{a}$ instead of $\hat{a}_{JC}(\vec{x}) $,
and $\hat{b}$ instead of $\hat{a}_{JC}(\vec{y}) $,
because we do not need the coordinate.
  In the superconducting qubit we have only two states 
  $|n_Q^*\rangle , |n_Q^*+1\rangle$ and the local operator $\hat{\phi}(\vec{x})$
  changes $|n_Q^*+1\rangle $ to $ |n_Q^*\rangle$.
  \begin{eqnarray}
\langle n_Q^*|\hat{\phi}(\vec{x})|n_Q^*+1\rangle=v \ , \nonumber\\
\langle n_Q^*+1|\hat{\phi}^\dagger(\vec{x})|n_Q^*\rangle=v
 \ \ .\label{Vc4e} 
 \end{eqnarray}
 Her we assume that $v$ is real.

 When we employ Pauli matrices for the down and up states
  $|n_Q^*\rangle$ and  $|n_Q^*+1\rangle$,
 the interaction (\ref{Vc4d}) can be expressed by $\sigma^{\pm}$, which is given by
  \begin{eqnarray}
 \hat{V}_{cQED}=g(\hat{a}\sigma^+ + \hat{a}^\dagger\sigma^-)+g(\hat{b}\sigma^+ + \hat{b}^\dagger\sigma^-)
  \ \ .\label{Vc4f}
\end{eqnarray}
 Here we use $g=g_{p}v$.
  If we measure the position-independent correlation by this interaction  $\hat{V}_{cQED}$ in experiments,
  we can conclude that the cluster property is violated.
  We will consider Hamiltonian that contains $\hat{V}_{cQED}$ in detail in next section.


%
\vskip 1cm

%

\section{Jaynes-Cummings Model on Two Cavities}
We will consider  Jaynes-Cummings model\cite{JC1}\cite{JC2} of  photons on two cavities  and
the two levels qubits, which are described by the spin $1/2$ states.
The annihilation operator of photon at cavity A is denoted by $\hat{a}$,
while that at cavity B is done by $\hat{b}$.
As shown in section 2, we defined  Hamiltonian by
\begin{eqnarray}
 \hat{H}= \omega_a\hat{a}^\dagger \hat{a} +\omega_b \hat{b}^\dagger \hat{b}+ \frac{\Omega}{2}\sigma^z\nonumber \\
+g( \hat{a}^\dagger \sigma^- + \hat{b}^\dagger \sigma^-
+\hat{a}\sigma^+ + \hat{b}\sigma^+)
\ \ .\label{M1a}
\end{eqnarray}
Here $\hat{a}$, $\hat{b}$ and $\sigma^\alpha$ are defined by
\begin{eqnarray}
 [\hat{a}\ , \  \hat{a}^\dagger ]=1 \ \ , \ \  [\hat{b}\ , \  \hat{b}^\dagger ]=1
 \ \ , \nonumber \\
 \ \  [\hat{a}\ , \  \hat{b}]=0 \ \ , \ \  [\hat{a}\ , \  \hat{b}^\dagger ]=0 \ \ ,
 \nonumber \\
\hskip 0.3cm [\sigma^x\ , \  \sigma^y ] =2i \sigma^z\ \ , \ \ \sigma^{\pm}\equiv \frac{1}{2}(\sigma^x \pm i \sigma^y)
\ \ .\label{M1a0}
\end{eqnarray}

$\hat{N}_{ab}=\hat{a}^\dagger \hat{a}+ \hat{b}^\dagger \hat{b}+\sigma^z/2$
commutes with Hamiltonian (\ref{M1a}), by which
    eigenvalue $N_{ab}$  of $\hat{N}_{ab}$ is fixed for the energy eigenstate.

In the superconducting qubit   coupling $g$ is much smaller than
 photon frequency  $\omega_a$ or  $\omega_b$. 
In the experiments by cQED, two cases have been examined extensively\cite{cQED2}.
The first case is that $\omega_a$ and  $\omega_b$ equal to  energy gap $\Omega$. In the second case  $g$ is smaller than the difference $\Omega- \omega_{a(b)}$.
\vskip 0.5cm
\subsection{ $\Omega=\omega_a=\omega_b$}
Next, we consider a case that $\omega_a=\omega_b=\Omega$,
where they are denoted by $\omega$.
In this case 
we introduce new boson operators $\hat{a}_\pm$ instead of $ \hat{a},\hat{b}$.
Then  Hamiltonian is replaced by
\begin{eqnarray}
\hat{a}_\pm\equiv \frac{1}{\sqrt{2}}( \hat{a}\pm\hat{b})
\ \ , \ \ [ \hat{a}_\pm \ , \ \hat{a}_\pm^\dagger]=1
\ \ , \nonumber\\
 \hat{H}
 = \omega(\hat{a}_+^\dagger \hat{a}_+ + \hat{a}_-^\dagger \hat{a}_-)+ \frac{\omega}{2}\sigma^z+\nonumber\\
+\sqrt{2}g( \hat{a}_+^\dagger \sigma^- 
+\hat{a}_+\sigma^+ ) \ \ , \nonumber\\
\hat{N}_{ab}=\hat{a}_+^\dagger \hat{a}_+ + \hat{a}_-^\dagger \hat{a}_-+\sigma^z/2
\ \ .\label{M1aa} 
\end{eqnarray}
On   $ \hat{a}_+$ and $ \sigma^\alpha$, this Hamiltonian is
equivalent to Jaynes-Cummings model of one photon so that
 we obtain  the eigenstates and
the energy  eigenvalues  easily, as described in appendix B.
The basis state is given  by
$|\downarrow,n_a,n_b\rangle $ and $|\uparrow,n_a,n_b\rangle $.
The lowest energy  state of the qubits is denoted by $|\downarrow\rangle $,
while the excited state is done by $|\uparrow\rangle $.
$|n_{a(b)}\rangle $ is  the state of eigenvalue $n_{a(b)}$
 of  number operator $\hat{n}_{a(b)}=\hat{a}_{a(b)}^\dagger
 \hat{a}_{a(b)}$.
We define them explicitly by
\begin{eqnarray}
\sigma^z|\downarrow,n_a,n_b\rangle=- |\downarrow,n_a,n_b\rangle \ \ ,\nonumber\\
\ \ 
\sigma^z|\uparrow,n_a,n_b\rangle=|\uparrow,n_a,n_b\rangle
\ \  , \nonumber \\
\hat{n}_a|*,n_a,n_b\rangle= |*,n_a,n_b\rangle n_a \ \ , \nonumber\\\ \ 
\hat{n}_b|*,n_a,n_b\rangle= |*,n_a,n_b\rangle n_b \ \ , \ \ 
 (*=\downarrow,\uparrow)
\ \ .\label{M1ab}
\end{eqnarray}
Here we focus our interest to the states whose the photon number is $0$ or $1$.
The lowest energy state of Hamiltonian (\ref{M1aa}) is given by 
\begin{eqnarray}
E_{0}=- \frac{\omega}{2} \ \ , \ \ |E_0\rangle= |\downarrow,0_a,0_b\rangle
\ \ .\label{M1ac}
\end{eqnarray}

We show three excited states which are given by
\begin{eqnarray}
|\psi_1\rangle\equiv |\psi(-,\frac{1}{2},0)\rangle= \hskip 1cm \nonumber\\
=\frac{-1}{\sqrt{2}}|\uparrow,0_a,0_b\rangle
+\frac{1}{2}(|\downarrow,1_a,0_b\rangle+ |\downarrow,0_a,1_b\rangle)
\ \  , \nonumber \\
|\psi_2\rangle\equiv |\psi(-,\frac{1}{2},1)\rangle
= \frac{1}{\sqrt{2}}(|\downarrow,1_a,0_b\rangle- |\downarrow,0_a,1_b\rangle)
\ \  , \nonumber \\
|\psi_3\rangle\equiv |\psi(+,\frac{1}{2},0)\rangle= \hskip 1cm\nonumber\\
=\frac{1}{\sqrt{2}}|\uparrow,0_a,0_b\rangle
+\frac{1}{2}(|\downarrow,1_a,0_b\rangle+ |\downarrow,0_a,1_b\rangle)
\ \ .\label{M1ad}
\end{eqnarray}
 The energy eigenvalues are given by
 \begin{eqnarray}
E_1-E_0=\omega- g\sqrt{2}\ \ , \ \ E_2-E_0=\omega \ \ , \nonumber\\
 \ \ E_3-E_0=\omega+ g\sqrt{2}
\ \ .\label{M1ae}
\end{eqnarray}
If  we observe  the three energy gaps in the experiment of the spectroscopy, 
it strongly supports our arguments on the cluster property.
\vskip 0.5cm 
 \subsection{ $g\ll \Omega- \omega_{a(b)}$}
 We will consider another  case that $g\ll  \Omega- \omega_{a(b)}$. Then
we obtain the effective Hamiltonian  $H_{eff}$ in order of $g^2$.
 See the  calculations given in appendix C.
\begin{eqnarray}
\hat{H}_{eff} =\frac{1}{2}(\Omega+g^2\gamma_{ab})\sigma^z
+\frac{g^2 \gamma_{ab}}{2}\hskip 2cm
\nonumber\\
+( \omega_a +g^2\gamma_a\sigma^z)  \hat{a}^\dagger  \hat{a}
+( \omega_b +g^2\gamma_b\sigma^z)  \hat{b}^\dagger \hat{b}\nonumber\\
+\frac{g^2\gamma_{ab} }{2}( 
 \hat{a}^\dagger  \hat{b}+  \hat{b}^\dagger \hat{a}) \sigma^z
\ \ .\label{M1b}
\end{eqnarray}
Here 
\begin{eqnarray}
\gamma_a \equiv \frac{1}{\Omega-  \omega_a} \ \ , \ \ 
\gamma_b\equiv \frac{1}{\Omega-  \omega_b} \ \ , \ \ 
\gamma_{ab}\equiv \gamma_{a}+\gamma_{b}
\ \ .\label{M1c}
\end{eqnarray}
 In the effective Hamiltonian (\ref{M1b}), we have only one operator
 $\sigma^z$ on the superconducting state so that
 the qubit state does not change by the interaction with photons.
 Therefore, we  make $\sigma^z$ the constant
 $\pm 1$ to examine Hamiltonian  (\ref{M1b}). Then
 we introduce the parameters $\omega_{a,z}$ 
 $\omega_{b,z}$ and  $g_{2,z}$.
 \begin{eqnarray}
\omega_{a,z}\equiv \omega_a+g^2\gamma_a\sigma^z \ \  ,\ \ 
\omega_{b,z}\equiv \omega_b+g^2\gamma_b\sigma^z \ \  , \nonumber\\
g_{2,z}\equiv \frac{g^2\gamma_{ab} }{2}\sigma^z
\ \ , \nonumber\\
C_z \equiv  \frac{1}{2}\{\Omega+g^2 \gamma_{ab})\} \sigma^z
+\frac{1}{2}g^2 \gamma_{ab}
\ \ .\label{M1d}
\end{eqnarray}
Therefore, Hamiltonian (\ref{M1b}) is expressed by
the following.
 \begin{eqnarray}
\hat{H}_{eff} =C_z
+\omega_{a,z} \hat{a}^\dagger  \hat{a}
+ \omega_{b,z}  \hat{b}^\dagger \hat{b} +\nonumber\\
+g_{2,z}( \hat{a}^\dagger  \hat{b}+  \hat{b}^\dagger \hat{a}) \ \ 
\ \ .\label{M1e}
\end{eqnarray}

In  cQED of one cavity for  superconducting qubit,
the frequency of the photon depends on the qubit state.
This feature is found in effective  Hamiltonian (\ref{M1b}), where the frequency of photon
$\omega_{a,z}$ at cavity A  for the lowest energy   state differs from that for
 the excited one.
Therefore,  we can  observe the change of the frequency due to the state.

We will discuss  characteristic features of   the effective Hamiltonian (\ref{M1e}) with two  cavities.
For this discussion
 we express  $ \hat{H}_{eff} $ by
new operators $\hat{\alpha},\hat{\beta}$.
\begin{eqnarray}
\hat{a}= \cos\theta \hat{\alpha} +\sin\theta \hat{\beta}
\ \ , \ \
\hat{b}= -\sin\theta \hat{\alpha} +\cos\theta \hat{\beta}
\ \ .\label{M2a}
\end{eqnarray}
To  eliminate the term of $\hat{\alpha}^\dagger \hat{\beta}$ in $H_{eff} $ ,
we determine the angle $\theta$ by
 $$\tan(2\theta)=\frac{2g_{2,z}}{   \omega_{bz}-\omega_{az}}     \ \ . $$
 Therefore, we obtain
\begin{eqnarray}
 \hat{H}_{eff} 
=\omega_{\alpha} \hat{\alpha}^\dagger \hat{\alpha} 
+\omega_{\beta}  \hat{\beta}^\dagger \hat{\beta}+C_z 
\ \ .\label{M2c}
\end{eqnarray}
Here 
\begin{eqnarray}
  \omega_\alpha=\frac{\omega_{az}+\omega_{bz}}{2}
- \frac{1 }{2}\sqrt{(\omega_{bz}-\omega_{az})^2+4g_z^2}
   \ \ , \nonumber\\
   \omega_\beta=\frac{ \omega_{az}+\omega_{bz}}{2}
+ \frac{1 }{2}\sqrt{(\omega_{bz}-\omega_{az})^2+4g_{2,z}^2}   
\ \ .\label{M2d}
\end{eqnarray}

To observe the energy eigenvalues,  we will consider the time evolution of the photon operator, where we have
\begin{eqnarray}
 \hat{a}(t)=e^{i\hat{H}_{eff} t}\hat{a}(0) e^{-i\hat{H}_{eff} t} \ \ , 
\nonumber\\
\hat{b}(t)=e^{i\hat{H}_{eff} t}\hat{b}(0) e^{-i\hat{H}_{eff} t} \ \ .
\end{eqnarray}
We calculate $ \hat{a}(t) $ and $\hat{b}(t)$ explicitly.
\begin{eqnarray}
\hat{a}(t)
=\cos\theta e^{-i\omega_\alpha t}\{ \cos\theta \hat{a}(0) -\sin\theta \hat{b}(0)\}
+\nonumber\\
+\sin\theta e^{-i\omega_\beta t} \{ \sin\theta \hat{a}(0) +\cos\theta \hat{b}(0)\}
\ \ , \nonumber\\
\hat{b}(t)
=-\sin\theta e^{-i\omega_\alpha t}\{ \cos\theta \hat{a}(0) -\sin\theta \hat{b}(0)\}
+\nonumber\\
+\cos\theta e^{-i\omega_\beta t} \{ \sin\theta \hat{a}(0) +\cos\theta \hat{b}(0)\}
\ \ .\label{Ap5a}
\end{eqnarray}
For concrete examples
we calculate the expectation of  photon number  operator
 $\hat{a}(t)^\dagger\hat{a}(t)$ in cavity A.
For  initial state $|\Psi\rangle $,
we have the expectation values of the number operators.
\begin{eqnarray}
\langle \Psi |   \hat{a}^\dagger(0) \hat{a}(0) |\Psi\rangle =n_a  \ \ , \nonumber\\
\langle \Psi |   \hat{b}^\dagger(0) \hat{b}(0) |\Psi\rangle =n_b \ \ , \nonumber\\
\langle \Psi |   \hat{a}^\dagger(0) \hat{b}(0) |\Psi\rangle =n_{ab}
\ \ .\label{Hf2a}
\end{eqnarray}
Therefore, we obtain
  \begin{eqnarray}
  \langle   \Psi |\hat{a}^\dagger(t)\hat{a}(t)|  |\Psi\rangle\hskip 2cm \nonumber\\
 =(\cos^4\theta+\sin^4\theta)n_a+2\cos^2\theta\sin^2\theta n_b
  +\nonumber\\
 +2(-\cos^2\theta+\sin^2\theta)\cos\theta\sin\theta (n_{ab} +n_{ab}^*)+
 \nonumber\\
+ \{2\cos^2\theta \sin^2\theta(n_a-n_b) + \hskip 1cm\nonumber\\
+\cos\theta \sin\theta (\cos^2\theta -\sin^2\theta )(n_{ab} +n_{ab}^*)  \}
\times \nonumber\\
 \times \cos\{( \omega_\alpha-\omega_\beta)t\}  +\hskip 2cm\nonumber\\
 +i\cos\theta \sin\theta (n_{ab} -n_{ab}^*)  
\sin\{( \omega_\alpha-\omega_\beta)t\}
\ \ .\label{Ap6d}
\end{eqnarray}

For numerical discussions, we assume
 the initial state is the photon state with the fixed number, i.e., $|\Psi\rangle =|n_a\rangle_a|n_b\rangle_b $,  i.e., $n_{ab}=0$ for  simplicity.
In this case  the time evolution is given by
  \begin{eqnarray}
\langle \Psi |\hat{a}^\dagger(t)\hat{a}(t)  |\Psi\rangle= \hskip 1cm\nonumber\\
 =(\cos^4\theta+\sin^4\theta)n_a+2\cos^2\theta\sin^2\theta n_b+
 \nonumber\\
+ 2\cos^2\theta \sin^2\theta(n_a-n_b) 
 \cos\{( \omega_\alpha-\omega_\beta)t\} 
\ \ .\label{Fg1aa}
\end{eqnarray}

In order to observe the oscillation, the magnitude of the second term in (\ref{Fg1aa}) is required to be comparable with that of the constant term.
Therefore, we will consider a ratio $r_a$ of the magnitude of  the oscillation term to
that of the constant term.
  \begin{eqnarray}
  r_a\equiv \frac{2\cos^2\theta \sin^2\theta(n_a-n_b) }{ 
  (\cos^4\theta+\sin^4\theta)n_a+ 2\cos^2\theta \sin^2\theta n_b}
   \nonumber\\
   =\frac{2g_{2,z}^2(n_a-n_b) }{ 
  \{(\omega_{a,z}-\omega_{b,z})^2+2g_{2,z}^2\}n_a+2g_{2,z}^2 n_b}
  \nonumber\\
   =\frac{(1-n_b/n_a) }{ 
  (\omega_{a,z}-\omega_{b,z})^2/(2g_{2,z}^2)+ 1+ n_b/n_a}
\ \ .\label{Fg1b}
\end{eqnarray}
This ratio becomes the maximum value of $1$, when $\omega_{a,z}=\omega_{b,z}$ and $n_b=0$.
Note that it increases when $(\omega_{a,z}-\omega_{b,z})/g_{2,z}$ becomes small
for any $n_b \le n_a$.
In Figure 1 we plot the ratio $r_a$ as a function of 
$(\omega_{a,z}-\omega_{b,z})/g_{2,z}$.
We find that the ratio is greater than $0.1$ when $n_b/n_a $ is less than $0.8$
for $ \omega_{a,z}=\omega_{b,z}$, i.e. $ \omega_a=\omega_b$.

In the above  case we need that $|(\omega_{a,z}-\omega_{b,z})/g_{2,z}|$ is
smaller than $1$ to observe the oscillation.
We confirm whether this requirement can be realized for experiments.
Here we assume that in experiments we   can fix the frequency of the photon
with the precision $\Delta \omega$.
Therefore, we can make $|\omega_{a}-\omega_{b}|$ to be of order of
the precision $\sqrt{2}\Delta \omega$.
Since $\omega_{a}\sim \omega_{b}$, we replace them by a frequency
$\omega$.
By this replacement,  we have  $|g_{2,z}|\sim g^2/(\Omega-\omega)$.
Therefore, the requirement of  $|(\omega_{a,z}-\omega_{b,z})/g_{2,z}| < 1$ 
is equivalent to
  \begin{eqnarray}
\sqrt{2}\Delta \omega < \frac{g^2}{(\Omega-\omega)}
\ \ .\label{Fg1bb}
\end{eqnarray}

Using  the values suggested in the work\cite{cQED2},
we examine the condition (\ref{Fg1bb}).
They are given by
$$ g/\omega=5\times 10^{-3} \ \ , \ \ \frac{(\Omega-\omega)}{\omega}=0.1 \ \ ,  \ \ \Delta \omega=10^{-4}\omega \ \ 
.$$
Therefore, we have
  \begin{eqnarray}
\frac{\sqrt{2}\Delta \omega }{ \omega }\sim 1.4\times 10^{-4} <
\frac{g^2}{(\Omega-\omega)\omega} = 2.5\times 10^{-4}
\ \ .\label{Fg1bc}
\end{eqnarray}
The condition  (\ref{Fg1bb}) is satisfied.
The extensive study\cite{cQED2} is made  in order to realize cQED for
the superconducting qubit in experiments, where the   conservative value
is used for $g$.  
The experiment\cite{cQED6} 
 reported that  $g/\omega=1.8\times 10^{-2}$ and $\Delta \omega=2.3\times 10^{-4}\omega$. In this case we find the following relation for
 $(\Omega-\omega)/\omega=0.1$.
   \begin{eqnarray}
\frac{\sqrt{2}\Delta \omega }{ \omega }\sim 3.2\times 10^{-4} <
\frac{g^2}{(\Omega-\omega)\omega} = 3.2\times 10^{-3}
\ \ .\label{Fg1bd}
\end{eqnarray}
Therefore, we have enough precision for the observation.
Also,  the experiment\cite{cQED7} used 
 values  of $g/\omega=2.2\times 10^{-2}$ and $\Delta \omega=0.77\times 10^{-4}\omega$, so that  for $(\Omega-\omega)/\omega=0.1$. we find that
   \begin{eqnarray}
\frac{\sqrt{2}\Delta \omega }{ \omega }\sim 1.1\times 10^{-4} <
\frac{g^2}{(\Omega-\omega)\omega} = 4.8\times 10^{-3}
\ \ .\label{Fg1be}
\end{eqnarray}
By these results  we conclude that  the experiments can satisfy our requirement $|\omega_a-\omega_b| < |g_{2,z}|$.
\vskip 0.5cm
\subsection{Noise in Time Evolution of Photon Number}
We will discuss negative effects by the noise in experiments
of the time evolution of photon number $\langle \Psi |\hat{a}^\dagger(t)\hat{a}(t)  |\Psi\rangle $ in (\ref{Fg1aa}), which has been studied in a previous
subsection.
Here let us consider 
 a case that theoretically  we expect $\omega_a=\omega_b$, but
the difference $\omega_a-\omega_b$ contains  the Gauss noise
 $\delta\omega $.
  The probability function $p(\delta\omega) $  for $ \delta\omega$ is the following.
  \begin{eqnarray}
  \omega_a-\omega_b=\delta\omega \ \ , \nonumber \\
  p(\delta\omega)= \frac{1}{\sqrt{2\pi}\sigma}
  e^{-\frac{\delta\omega^2}{2\sigma^2}}
\ \ .\label{Fg1c}
\end{eqnarray}
The effects of $\delta\omega$ are found in both the amplitude $ 2\cos^2\theta\sin^2\theta$ and phase $\cos\{( \omega_\alpha-\omega_\beta)t\}$ of the oscillation.
The effect on the amplitude does not depend on the time,
while the effect on the phase becomes large as the time increases.
Therefore, we consider the effect on the phase only.
The energy difference $\omega_\alpha-\omega_\beta$ is a function of
 $\delta\omega$.
 \begin{eqnarray}
 \omega_\alpha -\omega_\beta=\delta\omega+
 \sqrt{( \delta\omega)^2+4g_{2,z}^2}
 \ \ .\label{Fg1cc}
\end{eqnarray}
 The probability expectation of $\cos\{( \omega_\alpha-\omega_\beta)t\}$
 is defined by
 \begin{eqnarray}
 <\cos\{( \omega_\alpha-\omega_\beta)t\}>  \equiv \hskip 1cm \nonumber\\
\equiv \int_{-\infty}^{\infty} p(\delta\omega) \cos\{( \omega_\alpha-\omega_\beta)t\}  \nonumber\\
 =\frac{1}{2}\int_{-\infty}^{\infty} p(\delta\omega)\{
  e^{ i\delta\omega t+i\sqrt{\delta\omega^2+4g_{2,z}} t}+c.c.\}
  d \delta\omega 
 \ \ .\label{Fg1d}
\end{eqnarray}
Since the variance $\sigma^2$ is smaller than $g_{2,z}^2$,
we approximate the squared root in the phase by
 \begin{eqnarray}
 e^{ i\delta\omega t+i\sqrt{\delta\omega^2+4g_{2,z}^2} t} \sim
  \nonumber\\
 \sim  \exp\{ i\delta\omega t+
i(2g_{2,z}t+ \frac{\delta\omega^2}{4g_{2,z}} t)\}
 \ \ .\label{Fg1da}
\end{eqnarray}
Then we carry out the integral (\ref{Fg1d}). The result is given by
 \begin{eqnarray}
 <\cos\{( \omega_\alpha-\omega_\beta)t\} >=\hskip 1cm  \nonumber\\
  = \exp(-\frac{t^2 \cos\eta}{4A_0} )   \frac{1}{\sqrt{2A_0}\sigma}
  \times  \ \ \ \ \nonumber\\
\times   \cos (2g_{2,z}t +\frac{t^2 \sin\eta}{4 A_0}- \frac{\eta}{2})\ \ , \nonumber\\
  A_0=\frac{1}{2\sigma^2}\{1+(\frac{\sigma^2 t}{2g_{2,z}})^2\}^{1/2}
  \ \ , \ \ \tan\eta = \frac{-\sigma^2 t}{2g_{2,z}}
\ \ .\label{Fg1e}
\end{eqnarray}

We plot $ <\cos\{( \omega_\alpha-\omega_\beta)t\} > $ in Figure 2
for various values of the ratio $\sigma/g_{2,z}$.
For $\sigma/g_{2,z}=0.2$ we find clear oscillation during two periods.
When $\sigma/g_{2,z}$ is larger than $0.3$, we can not neglect the dumping of the oscillation.
At $\sigma/g_{2,z}=0.5$, it is difficult to observe the oscillation during one period.
Therefore, we need that  the magnitude of the noise $\sigma$ is less than
$0.3g_{2,z}$.
This is not satisfied  for the parameters given by  (\ref{Fg1bc}),
while
our requirement is reasonable for those by (\ref{Fg1bd}) and (\ref{Fg1be}).

\begin{figure}[ht]%
\begin{center}
\scalebox{0.3}{\includegraphics{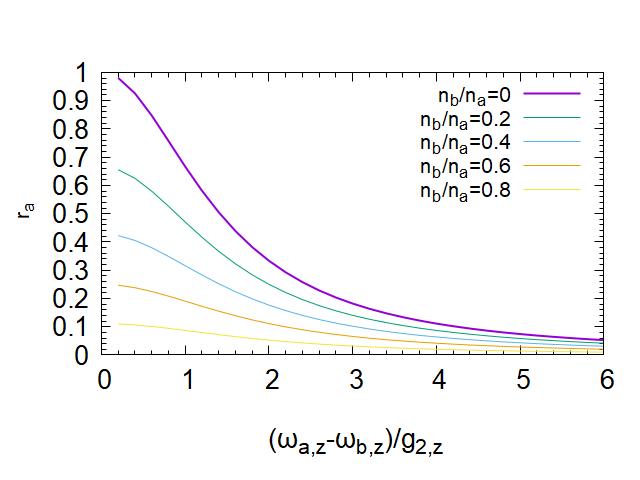}}
\caption{ Ratio $r_a$ of the oscillation term to the constant term,
which is defined in (\ref{Fg1b}).
The horizon axis is denoted by $(  \omega_{a,z}-\omega_{b,z})/g_{2,z}$.
The several curves are calculated for $n_b/n_a=0,0.2,0.4, 0.6 $
and $0.8$.
 }
\label{fig1}
\end{center}
\end{figure}

\begin{figure}[ht]%
\begin{center}
\scalebox{0.3}{\includegraphics{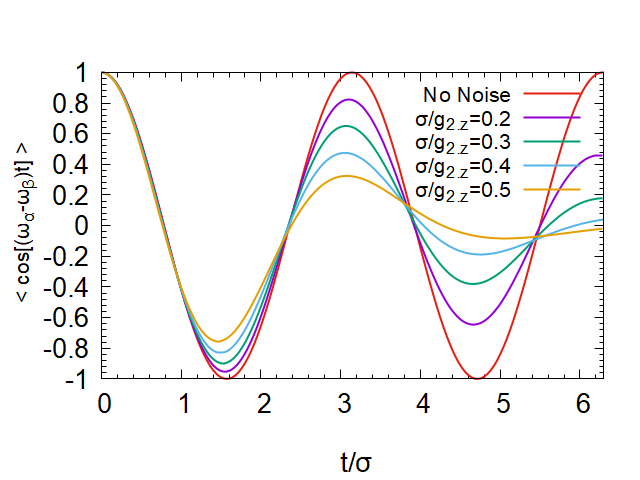}}
\caption{
$Y=<\cos\{( \omega_\alpha-\omega_\beta)t\} >$
which is defined in (\ref{Fg1e}).
The horizon axis is denoted by $t/\sigma$.
The several curves are calculated for $\sigma/g_{2,z}=0,0.2,0.3,0,4 $
and $0.5$.
 }
\label{fig2}
\end{center}
\end{figure}


\vskip 1cm
\section{Summary and Discussion}
One of the  most important concepts in quantum physics is
spontaneously symmetry breaking (SSB) in  the condense matter physics\cite{SSB1} and particle physics\cite{SSB2}.
The  essential  features of SSB are
the energy degeneracy and the modification of the eigenstate by the local operator\cite{My1}.
If the lowest energy state is stable, 
 the modification of ground state    by the local operator
implies that the cluster property
violates\cite{CP1}\cite{CP2}\cite{CP3}\cite{CP4}\cite{CP5}\cite{CP6}.
Due to the degeneracy of the energy eigen states, 
 the lowest energy state is not  stable.
Therefore, we have to  introduce  the explicitly breaking interaction in order to make the state
stable. As  results  the eigenstate is coherent on  charge number $n_Q$.
For this coherent state  the violation of the cluster property becomes tinny effects\cite{My1}\cite{My2}\cite{My3}.

In  the superconducting  qubit
the experimental researchers have  included the charge square operator so that
the eigenstate of   definite charge number $n_Q$ has become stable\cite{qbit0}.
In these systems one can modify the lowest energy eigen state into
the excited state has $n_Q+1$.
Using two states we realize a quantum bit (qubit), by which
we have made active study 
\cite{qbitReview1}\cite{qbitReview2}\cite{qbitReview3}\cite{qbitReview4}
for  quantum computing\cite{QC1}.

In this paper we have pointed out that for the superconducting qubit
 the cluster property is violated without any doubt. 
Then
we  proposed the method by cavity quantum electrodynamics with two cavities in order to
observe  the violation.
By extending  Jaynes-Cumming model\cite{JC1}\cite{JC2}, we have shown that
two photons in the  two cavities  correlate at  finite magnitude even if
these cavities are separated at the far distance.
%
%

In this work we discussed the cluster property on the charge qubit\cite{qbit0},
because we formed Cooper pair box, the quantum system with the definite charge number.
We will touch upon the possibility of experiments of two cavities in the charge qubit. 
The two or multi cavities have been studied in the theoretical point of view\cite{2cQED1}
as well as the experimental one\cite{2cQED2}.
The former researched them to control the strength of qubit-qubit coupling.
While the latter studied  the spurious mode to suppress
the decoherence of the qubit.
Therefore, we can argue that the experiment for our proposal method in this paper is possible.
More quantitative discussion is shown for brief.
For the study of the cluster property  we need to connect two cavities to the charge qubit without
 no interference between the photons in these cavities.
The key is the size of Cooper pair-box\cite{CB1}\cite{CB2}.
The work \cite{qbit0} reported that the size is $ 700  \times 50  \times 15  {\rm nm}^3 $,
while  the size  $30\times 110 \times 2260 {\rm nm}^3$  is mentioned in \cite{CB3}.
 The cavity cQED has  the size $10^{-6}\times 10^{-2} {\rm m}^2$ 
for the micro wave  of the frequency  10GHz.
Therefore, it becomes possible to realize   experiments for our proposal.
Note that the decoherence time, which is longer than $10 \mu {\rm sec}$\cite{qbitReview4},
 is enough long compared with the oscillation time $0.1\mu {\rm sec}$
 that is estimated in section 3.

Also, the long Cooper pair box will be promising candidates for them.
Therefore, the superconducting nanowire is quite interesting\cite{nanowire1}\cite{nanowire2}.
Specially  \cite{nanowire3}\cite{nanowire4} reported the realization of 
 Coulomb blockade in nanowire.

While we have discussed  the charge qubit\cite{qbit0},
we find the extensive studies on
the flux qubit\cite{qbitflux1}\cite{qbitflux2}
and the phase qubit\cite{qbitPhase1}\cite{qbitPhase2}\cite{qbitPhase3}.
It is quite interacting to study the cluster property in these types of
the superconducting qubit.
If there are  local operators or devices  that change the qubit state,
it is possible to detect
 the violation  for these systems

Final comments are made on  entanglement\cite{ET0}\cite{ET1}\cite{ET2}\cite{ET3}.
Entanglement   in the superconducting with cavity QED was discussed\cite{cQED5}.
In the proposed system of two cavities, the maximally entangled state is realized 
  for  two photons with the same frequency,
  as we pointed out it in section 3.
  Entanglement using   two cavities has been studied\cite{En2cQED1}\cite{En2cQED2},
  where the interaction between atoms and photons are investigated.
Entanglement of photons in the electric circuits using
two cavities will stimulate  the experimental researches because of
the  large freedom for designs of systems.




\vskip 1cm


\noindent
 {\large \bf Appendix A:
Indirect Observation for Violation of Cluster Property  }
\vskip 0.3cm
\noindent
In this appendix we describe an indirect method to observe the violation of cluster property (VCP).
Here we assume that  in spontaneous symmetry breaking (SSB)
the lowest energy state $ |E_{lst} \rangle $ with definite number of charges is stable
and  there exists local operator
$\hat{\phi}^\dagger(\vec{x})$ that can modify this state to the excited state
$ |E_{ex} \rangle $ with definite number of charges.
It implies that we have a finite value for $v$ which is defined by
$$v\equiv   \langle E_{ex}|\hat{\phi}^\dagger(\vec{x})|E_{lst} \rangle   \  \ .$$
Here we consider only these states in the further discussion.

Then we will consider the other degree of freedom $\hat{\psi}$
and  Hamiltonian $\hat{H}_\psi$ that is made of  $\hat{\psi}$ only.
By this assumption
the state is given by
\begin{eqnarray}
| \Psi\rangle =\sum_i | E_i \rangle |\psi_i \rangle c_i
\nonumber \\
=|E_{lst} \rangle|\psi\rangle c
+|E_{ex} \rangle|\psi'\rangle c'
  \ .\label{ApCa} 
\end{eqnarray}
Here  $ |\psi \rangle $ and $ |\psi' \rangle $ are  states on  $\hat{\psi}$.

In order to observe VCP, we introduce an interaction $\hat{H}_{ob}$.
\begin{eqnarray}
 \hat{H}_{ ob}\equiv g_{ ob}\{ \hat{O}(\vec{x})\hat{\phi}(\vec{x})
 + \hat{O}(\vec{y})\hat{\phi}(\vec{y})+c.c.\}
  \ .\label{ApC1b} 
\end{eqnarray}
Here  $ \hat{O}(\vec{x})$  is relevant on $\hat{\psi}$.
Also we assume $|g_{ob}| \ll 1$ so that
we can apply perturbation theory to examine the effect by $\hat{H}_{ob}$.
As results we obtain the effective Hamiltonian of $\hat{H}_{ob}$
on $\hat{\psi}$\cite{My3}.

\begin{eqnarray}
 \hat{H}_{ ob,eff}\equiv  \hskip 3cm \nonumber\\
\equiv \frac{1}{ E_{lst} -E_{ex} } \langle E_{lst} |\hat{H}_{  ob} |E_{ex} \rangle
  \langle E_{ex} |\hat{H}_{ ob} |E_{lst} \rangle \nonumber \\
  =
   \frac{1}{ E_{lst} -E_{ex} }\times   \hskip 2cm   \nonumber \\
  \times   \langle E_{lst} |
   g_{ ob}\{  \hat{O}(\vec{x})\hat{\phi}(\vec{x})
 + \hat{O}^\dagger(\vec{y})\hat{\phi}(\vec{y})\}  |E_{ex} \rangle \times 
  \nonumber \\
 \times \langle E_{ex} |
   g_{ob}\{  \hat{O}^\dagger(\vec{x})\hat{\phi}^\dagger(\vec{x})
 + \hat{O}(\vec{y})\hat{\phi}^\dagger(\vec{y})\}  |E_{lst} \rangle  \} \nonumber \\
 = \frac{g_{ob}^2}{ E_{lst} -E_{ex} }\times \hskip 2cm\nonumber \\
  \times \{  \hat{O}(\vec{x}) \hat{O}^\dagger(\vec{x})
 \langle E_{lst} |\hat{\phi}(\vec{x})  |E_{ex} \rangle   \langle E_{ex} |     \hat{\phi}^\dagger(\vec{x}) |E_{ex} \rangle + \nonumber \\
 + \hat{O}(\vec{y}) \hat{O}^\dagger(\vec{y})
 \langle E_{lst} |\hat{\phi}(\vec{y})  |E_{ex} \rangle   \langle E_{ex} |     \hat{\phi}^\dagger(\vec{y}) |E_{ex} \rangle+  \nonumber \\
+ \hat{O}(\vec{x}) \hat{O}^\dagger(\vec{y})
 \langle E_{lst} |\hat{\phi}(\vec{x})  |E_{ex} \rangle   \langle E_{ex} |     \hat{\phi}^\dagger(\vec{y}) |E_{ex} \rangle + \nonumber \\
+ \hat{O}(\vec{y}) \hat{O}^\dagger(\vec{x})
 \langle E_{lst} |\hat{\phi}(\vec{y})  |E_{ex} \rangle   \langle E_{ex} |     \hat{\phi}^\dagger(\vec{x}) |E_{ex} \rangle  \}
  \ .\label{ApC1c} 
\end{eqnarray}

Since  $  \langle E_{lst} |     \hat{\phi}(\vec{x}) |E_{ex} \rangle =v^* $,
we obtain
 \begin{eqnarray}
  \hat{H}_{ ob,eff}
  = \frac{g_{ ob}^2|v|^2}{ E_{lst} -E_{ex} } \{  \hat{O}(\vec{x}) \hat{O}^\dagger(\vec{x})+ \nonumber \\
+ \hat{O}(\vec{y}) \hat{O}^\dagger(\vec{y})
+ \hat{O}(\vec{x}) \hat{O}^\dagger(\vec{y})
+ \hat{O}(\vec{y}) \hat{O}^\dagger(\vec{x})
 \}
  \ .\label{ApC1d} 
\end{eqnarray}
Even if the operator at $\vec{x}$ goes away from  $\vec{y}$,
the magnitude of the interaction between $\hat{O}^\dagger(\vec{x}) $ and $\hat{O}(\vec{y})$
remains  constant.
Therefore, we find the non-local interaction between these operators.
Note that the non-zero value of $v$ is essential  for its existence.
Conclusively we can observe VCP by the probe interaction
between $\hat{\phi}$ and other $\hat{\psi}$.

\vskip 0.5cm

\noindent
 {\large \bf 
Appendix B:
Solutions for Jaynes-Cumming Model of  Photons with  Same Frequency}
\vskip 0.5cm
\noindent
We will consider a model of photons on two cavities and the two levels qubits,
which are described by the spin $1/2$ states.
These annihilation operators of two photons are denoted by $\hat{a},\hat{b}$, and 
a spin is expressed by Pauli matrix, $\sigma^\alpha$.
The basis states are  $|\downarrow,n_a,n_b\rangle $ and
$|\uparrow,n_a,n_b\rangle $, 
where $|\downarrow\rangle $ is the ground state on qubits.
and  $|\uparrow\rangle $  is the excited state.
 $n_a(n_b)$ is a number of photon $a(b)$.

 Here we suppose that two photons have the same frequency.
 We define Hamiltonian  given by
\begin{eqnarray}
 \hat{H}\equiv  \omega(\hat{a}^\dagger \hat{a} + \hat{b}^\dagger \hat{b})
 + \frac{\Omega}{2}\sigma^z+\nonumber \\
+g\{(\hat{a}^\dagger + \hat{b}^\dagger )\sigma^-
+(\hat{a} + \hat{b})\sigma^+)\}
\ \ .\label{Ap1a}
\end{eqnarray}
$ \hat{N}_{{\rm ab}}=\hat{a}^\dagger \hat{a}+  \hat{b}^\dagger  \hat{b}+\sigma^z/2$ commutes with $\hat{H}$ so that we can fix  eigen value $N_{{\rm ab}}$ of $\hat{N}_{{\rm ab}}$ in
the energy eigen state.

Since only  $(\hat{a} + \hat{b})$ couples with the spin operators,
 we introduce new boson operators $\hat{a}_\pm$ instead of $ \hat{a},\hat{b}$.
Then  Hamiltonian is changed by
\begin{eqnarray}
 \hat{H}= \omega(\hat{a}_+^\dagger \hat{a}_+ + \hat{a}_-^\dagger \hat{a}_-)+ \frac{\Omega}{2}\sigma^z+\nonumber \\
+\sqrt{2}g( \hat{a}_+^\dagger \sigma^- 
+\hat{a}_+\sigma^+ ) \ \ , \nonumber\\
\hat{N}=\hat{a}_+^\dagger \hat{a}_+ + \hat{a}_-^\dagger \hat{a}_-+\sigma^z/2
\ \ , \nonumber\\
\hat{a}_\pm\equiv \frac{1}{\sqrt{2}}( \hat{a}\pm\hat{b})
\ \ , \ \ [ \hat{a}_\pm \ , \ \hat{a}_\pm^\dagger]=1
\ \ .\label{Ap1b} 
\end{eqnarray}
%
Next we define  states $|\uparrow,n_+,n_-\rangle $ 
and $|\downarrow,n_+,n_-\rangle $ which are 
the eigenstates of   $\hat{a}_+^\dagger \hat{a}_+$ and
 $\hat{a}_-^\dagger \hat{a}_-$.
  \begin{eqnarray}
  |*,n_+=0,n_-=0\rangle =|*,n_a=0,n_b=0\rangle \ \ , \nonumber\\
 |*,n_+,n_-\rangle \equiv \nonumber \\
 (\hat{a}_{-}^\dagger )^{n_-} (\hat{a}_{+}^\dagger )^{n_+} |*,n_+=0,n_-=0\rangle C(n_+,n_-)\ \ , \nonumber\\
  C(n_+,n_-)=\frac{1}{\sqrt{n_+!n_-!}}\ \ , 
  \ \ ( *=\uparrow,\downarrow)\ \ ,
  \nonumber\\
  \hat{a}_+^\dagger \hat{a}_+|*,n_+,n_-\rangle=|*,n_+,n_-\rangle n_+ \ \ ,
  \nonumber\\
  \hat{a}_-^\dagger \hat{a}_-|*,n_+,n_-\rangle=|*,n_+,n_-\rangle n_- \ \ ,
   \nonumber\\
\sigma^z |\uparrow,n_+,n_-\rangle = |\uparrow,n_+,n_-\rangle 
\ \ , \nonumber \\
\ \ \sigma^z |\downarrow,n_+,n_-\rangle = |\downarrow,n_+,n_-\rangle (-1)
\ \ . \label{Ap1c}
\end{eqnarray}
Since $\hat{a}_{-}^\dagger\hat{a}_{-}$ commutes with $\hat{H} $ in (\ref{Ap1b}),
we fix $n_-$  as well as $N_{{\rm ab}}$ for the energy eigen state.
Therefore, we calculate the eigen state using   $ |\uparrow, N_{{\rm ab}}-n_--1/2,n_-\rangle$ and $ |\downarrow, N_{{\rm ab}}-n_-+1/2,n_-\rangle$.
\begin{eqnarray}
|\psi(+, N_{{\rm ab}},n_-)\rangle= \hskip 1cm\nonumber \\
=|\uparrow, N_{{\rm ab}}-n_--\frac{1}{2},n_-\rangle\cos\theta_{N_{{\rm ab}},n_-}
+\nonumber \\
+|\downarrow,  N_{{\rm ab}}-n_-+\frac{1}{2},n_-\rangle\sin\theta_{N_{{\rm ab}},n_-}
\ \  , \nonumber \\
|\psi(-, N_{{\rm ab}},n_-)\rangle = \hskip 1cm\nonumber \\
=|\uparrow,  N_{{\rm ab}}-n_--\frac{1}{2},n_-\rangle(-\sin\theta_{N_{{\rm ab}},n_-})
+\nonumber \\
+|\downarrow, N_{{\rm ab}}-n_-+\frac{1}{2},n_-\rangle\cos\theta_{N_{{\rm ab}},n_-}
\ \ .\label{Ap1dxx}
\end{eqnarray}
\begin{eqnarray}
\hat{H}|\psi(\pm, N_{{\rm ab}},n_-)\rangle =\hskip 1cm\nonumber \\
=|\psi(\pm, N_{{\rm ab}},n_-)\rangle E(\pm,N_{{\rm ab}},n_-)
\ \  , \nonumber \\
E(\pm,N_{{\rm ab}},n_-)= \omega N_{{\rm ab}}\pm \hskip 1cm \nonumber \\
\pm \frac{ \Omega -\omega}{2}
\sqrt{1+\frac{4g^2}{( \Omega -\omega)^2}( 2N_{{\rm ab}}-2n_- +1)} \ \  , \nonumber \\
\tan(2\theta_{N_{{\rm ab}},n_-})= \hskip 1cm\nonumber \\
=\frac{2g}{( \Omega-\omega)}\sqrt{2N_{{\rm ab}}-2n_- +1} \ \  , \nonumber \\
\hat{N}_{{\rm ab}}|\psi(\pm, N,n_-)\rangle =\hskip 1cm\nonumber \\
=|\psi(\pm, N_{{\rm ab}},n_-)\rangle N_{{\rm ab}}
\ \ .\label{Ap1d}
\end{eqnarray}
Here $N_{{\rm ab}}=n_-+\frac{1}{2},n_-+\frac{3}{2},n_-+\frac{5}{2},...$ and $n_-=0,1,2,...$ .
When $N_{{\rm ab}}=n_--\frac{1}{2}$ that is $n_+=0$,
$|\downarrow, N_{{\rm ab}}-n_-+\frac{1}{2},n_-\rangle=|\downarrow, 0,n_-\rangle$
cannot mix with another state.
Therefore, we have
\begin{eqnarray}
|\psi(-, n_--\frac{1}{2},n_-)\rangle =
| \downarrow, 0,n_-\rangle
\ \  , \nonumber \\
E(-,n_- -\frac{1}{2},n_-)= \omega n_- - \frac{\Omega}{2}
\ \ .\label{Ap1e}
\end{eqnarray}
 For $n_- =0$, we obtain 
 the lowest energy is $E_{0}=- \frac{\Omega}{2} $, 
 and its state is $|\downarrow,0,0\rangle$.

\vskip 0.5cm

\noindent
 {\large \bf 
Appendix C: Effective Hamiltonian for Small $g$}

\noindent
{ \bf C.1   Derivation of Effective Hamiltonian}
\vskip 0.3cm
\noindent
Hamiltonian of our model is given by
\begin{eqnarray}
\hat{H}=\hat{H}_0+\hat{V}\ \ ,  \nonumber\\
\hat{H}_0=\omega_a \hat{a}^\dagger\hat{ a}+\omega_b \hat{b}^\dagger \hat{b} +\frac{\Omega}{2} \sigma^z
\ \ , \nonumber\\
\hat{V}=g\{  (\hat{a}^\dagger+\hat{b}^\dagger )\sigma^-
+( \hat{a}+  \hat{b})\sigma^+\}
\ \ .\label{Ap2a}
\end{eqnarray}
In order to obtain the effective Hamiltonian for  small $g$,
we make the unitary transformation of Hamiltonian $\hat{H}_U\equiv
 \hat{U}\hat{H}\hat{U}^\dagger $,
where we have
\begin{eqnarray}
\hat{U} \equiv e^{\hat{A}}\ \ , \ \ 
\hat{A}^\dagger  =-\hat{A}
\ \ .\label{Ap2b}
\end{eqnarray}
Since $\hat{V}$ is order of $g$, we determine  $\hat{A}$ at the order of
$g$ to eliminate the term of $g$ in $\hat{H}_U$.
\begin{eqnarray}
\hat{U}\sim\hat{1}+\hat{A}+\frac{\hat{A}^2}{2}\ \ , \ \ 
\hat{U}^\dagger \sim \hat{1}-\hat{A}+\frac{\hat{A}^2}{2}\ \ , \nonumber\\
\hat{H}_U\sim(\hat{1}+\hat{A}+\frac{\hat{A}^2}{2})( \hat{H}_0+\hat{V})
(\hat{1}-\hat{A}+\frac{\hat{A}^2}{2})\nonumber\\
\sim \hat{H}_0+(-[\hat{H}_0,\ \hat{A}]+\hat{V})+\nonumber\\
+\{- \hat{A}\hat{H}_0\hat{A}+\frac{\hat{A}^2}{2} \hat{H}_0+\hat{H}_0\frac{\hat{A}^2}{2}-[\hat{V},\ \hat{A}]\}
\ \ .\label{Ap2c}
\end{eqnarray}
Note that
\begin{eqnarray}
\hat{A}^2\hat{H}_0=\hat{A}\hat{H}_0\hat{A}- \hat{A}[\hat{H}_0,\ \hat{A}]\ \ , 
\nonumber\\
\hat{H}_0\hat{A}^2=\hat{A}\hat{H}_0\hat{A}+ [\hat{H}_0,\ \hat{A}]\hat{A}\ \ .
\end{eqnarray}
In the last expression of (\ref{Ap2c}), we make the second term to vanish by  $\hat{A}$ that satisfies
the following condition.
$$ 
[\hat{H}_0,\ \hat{A}]=\hat{V} \ \ .
$$
Therefore, we have
\begin{eqnarray}
\hat{H}_U
\sim \hat{H}_0+
\{-\frac{1}{2} \hat{A}\hat{V}+\frac{1}{2} \hat{V}\hat{A}
-[\hat{V},\ \hat{A}]\}=\nonumber\\
=\hat{H}_0-\frac{1}{2}[\hat{V},\ \hat{A}]
\ \ .\label{Ap2d}
\end{eqnarray}

We will apply the above argument to our Hamiltonian (\ref{Ap2a}).
First, we assume the following expression for $ \hat{A}$. 
\begin{eqnarray}
 \hat{A}\equiv  \hat{A}_a+\hat{A}_b \ \ , \ \ \hat{A}_a \equiv  g \gamma_a( \hat{a}\sigma^+-\hat{a}^\dagger \sigma^-) 
\ \ , \nonumber\\ 
\hat{A}_b \equiv g \gamma_b( \hat{b}\sigma^+-\hat{b}^\dagger \sigma^-) 
\ \ .\label{Ap2e}
\end{eqnarray}
Here we introduce the decomposition of $\hat{V}$.
\begin{eqnarray}
\hat{V}\equiv  \hat{V}_a+\hat{V}_b \ \ ,\ \ 
\hat{V}_a \equiv  g( \hat{a}\sigma^++ \hat{a}^\dagger \sigma^-) 
\ \ , \nonumber\\ 
\hat{V}_b \equiv  g( \hat{b}\sigma^++ \hat{b}^\dagger \sigma^-) 
\ \ .\label{Ap2f}
\end{eqnarray}

Using these notations, we calculate $[\hat{H}_0,\ \hat{A}_a]$.
\begin{eqnarray}
[\hat{H}_0,\ \hat{A}_a]=g\gamma_a [\hat{H}_0,\hat{a}\sigma^+]
-g\gamma_a [\hat{H}_0,\hat{a}^\dagger\sigma^-]
 \ \ , \ \  \nonumber\\
\hskip 0.3cm [\hat{H}_0,\hat{a}\sigma^+]
=\omega_a[\hat{a}^\dagger \hat{a}, \hat{a}\sigma^+]+ \frac{\Omega}{2}[ \sigma^z,\hat{a}\sigma^+]=
 \nonumber\\
 =-(\omega_a-\Omega)\hat{a}\sigma^+
  \ \ , \hskip 1cm  \nonumber\\
  \hskip 0.3cm [\hat{H}_0,\hat{a}^\dagger\sigma^-]
 =(\omega_a-\Omega)\hat{a}^\dagger\sigma^-
\ \ , \ \  \nonumber\\
\hskip 0.3cm 
[H_0,A_a]=-g\gamma_a(\omega_a-\Omega)(\hat{a}\sigma^+ +  \hat{a}^\dagger \sigma^-)
\ \ .\label{Ap3a}
\end{eqnarray}
Since the last should be equal to $V_a$,
we have the following equations on $\gamma_a$,
and the solutions.
\begin{eqnarray}
-g\gamma_a(\omega_a-\Omega)=g\ \ , \ \
\gamma_a=\frac{1}{ \Omega-  \omega_a} 
\ \ .\label{Ap3b}
\end{eqnarray}
Also, we will make the same calculation for $\gamma_b$. 
\begin{eqnarray}
-g\gamma_b(\omega_b-\Omega)=g\ \ , \ \ 
\gamma_b=\frac{1}{ \Omega- \omega_b}
\ \ .\label{Ap3c}
\end{eqnarray}
We will calculate $[V,A]$
in (\ref{Ap2d}). First 
we should note that 
$$[\hat{V},\hat{A}]  =  [\hat{V}_a,\hat{A}_a]+ [\hat{V}_a,\hat{A}_b] +
[\hat{V}_b,\hat{A}_a] +[\hat{V}_b,\hat{A}_b] \ \ .$$
Then we calculate $   [\hat{V}_a,\hat{A}_a]$.
\begin{eqnarray}
[\hat{V}_a,\hat{A}_a]
=g^2 \gamma_a\{ [\hat{a}\sigma^+, -\hat{a}^\dagger \sigma^-]
+ [\hat{a}^\dagger \sigma^-, \hat{a}\sigma^+]\}=
\nonumber\\
=2g^2 \gamma_a\{ \hat{a}[\sigma^+, - \hat{a}^\dagger \sigma^-]
+[ \hat{a}, - \hat{a}^\dagger \sigma^-]\sigma^+
\} =\nonumber\\
=2g^2 \gamma_a\{ - \hat{a} \hat{a}^\dagger \sigma^z
-1\cdot \sigma^-\sigma^+\}=
\nonumber\\
=2g^2 \gamma_a\{ -( \hat{a}^\dagger  \hat{a}+1)\sigma^z
-\frac{1}{2}(1- \sigma^z)\} =\nonumber\\
= -2g^2 \gamma_a\{ \hat{a}^\dagger  \hat{a}\sigma^z+\frac{1}{2}(1+ \sigma^z)\}
\ \ .\label{Ap3d}
\end{eqnarray}
Therefore, we obtain
the sum of $ [\hat{V}_a,\hat{A}_a]$
 and 
$[\hat{V}_b,\hat{A}_b]$.
\begin{eqnarray}
 [\hat{V}_a,\hat{A}_a]+ [\hat{V}_b,\hat{A}_b]=\hskip 1cm\nonumber\\
= -2g^2 \gamma_a \hat{a}^\dagger \hat{a}\sigma^z
-2g^2 \gamma_b \hat{b}^\dagger \hat{b}\sigma^z-\nonumber\\
-g^2( \gamma_a+\gamma_b)(1+ \sigma^z)
\ \ .\label{Ap3e}
\end{eqnarray}
Also for $[\hat{V}_a,\hat{A}_b]$  we find  the following result.
\begin{eqnarray}
 [\hat{V}_a,\hat{A}_b]=\hskip 2cm\nonumber\\
=g^2 \gamma_b\{ [\hat{a}\sigma^+, -\hat{b}^\dagger \sigma^-]
+ [\hat{a}^\dagger \sigma^-, \hat{b}\sigma^+]\}=
\nonumber\\
=g^2 \gamma_b\{- \hat{a} \hat{b}^\dagger  [\sigma^+, \sigma^-]
+ \hat{a}^\dagger \hat{b} [\sigma^-, \sigma^+]\}
=\nonumber\\
=-g^2 \gamma_b( \hat{a} \hat{b}^\dagger + \hat{a}^\dagger \hat{b})\sigma^z
\ \ ,\label{Ap3f} \\
\hskip 0.1cm  [\hat{V}_a,\hat{A}_b] +[\hat{V}_b,\hat{A}_a] =  \hskip 1cm
\nonumber\\
= -g^2( \gamma_a+\gamma_b) 
( \hat{a} \hat{b}^\dagger + \hat{a}^\dagger \hat{b})
 \sigma^z
\ \ .\label{Ap3g}
\end{eqnarray}
Finally the result for  $[\hat{V},\hat{A}]/2$ is given by
\begin{eqnarray}
\frac{1}{2} [\hat{V},\hat{A}]=\frac{1}{2}\{-2g^2 \gamma_a\hat{a}^\dagger \hat{a}\sigma^z -\nonumber\\
-2g^2 \gamma_b  \hat{b}^\dagger \hat{b}\sigma^z
-g^2( \gamma_a+\gamma_b)(1+ \sigma^z)\}-\nonumber\\
-g^2\frac{1}{2}( \gamma_a+\gamma_b) 
( \hat{a} \hat{b}^\dagger + \hat{a}^\dagger \hat{b})\sigma^z
\ \ .\label{Ap3h}
\end{eqnarray}

We define  effective Hamiltonian $\hat{H}_{eff}$ as  $\hat{H}_U$ in (\ref{Ap2d}).
\begin{eqnarray}
\hat{H}_{eff}\equiv \hat{H}_0-\frac{1}{2}[\hat{V},\ \hat{A}] \hskip 1cm    \nonumber\\
=\frac{1}{2}\{\Omega+g^2( \gamma_a+\gamma_b)\} \sigma^z
+\frac{1}{2}g^2( \gamma_a+\gamma_b)+\nonumber\\
+( \omega_a +g^2\gamma_a\sigma^z) \hat{a}^\dagger \hat{a}
+( \omega_b +g^2\gamma_b\sigma^z) \hat{b}^\dagger \hat{b}+\nonumber\\
+\frac{1}{2}g^2( \gamma_a+\gamma_b) 
( \hat{a} \hat{b}^\dagger + \hat{a}^\dagger \hat{b})
\sigma^z
\ \ .\label{Ap3i}
\end{eqnarray}
Using $\gamma_{ab}\equiv \gamma_a+\gamma_b$,
we obtain the final expression for $ \hat{H}_{eff}$, which is  given by
\begin{eqnarray}
\hat{H}_{eff}
=\frac{1}{2}\{\Omega+g^2\gamma_{ab}\} \sigma^z
+\frac{1}{2}g^2\gamma_{ab}+\nonumber\\
+( \omega_a +g^2\gamma_a\sigma^z) \hat{a}^\dagger \hat{a}
+( \omega_b +g^2\gamma_b\sigma^z) \hat{b}^\dagger \hat{b}+\nonumber\\
+\frac{1}{2}g^2 \gamma_{ab}
( \hat{a} \hat{b}^\dagger + \hat{a}^\dagger \hat{b})
\sigma^z
\ \ .\label{Ap3j}
\end{eqnarray}

\vskip 0.5cm

\noindent
{ \bf C.2 
 Eigen Value of Effective Hamiltonian}
 \vskip 0.3cm
 \noindent
The effective Hamiltonian (\ref{Ap3i}) is diagonal on the spin eigen value
$\pm 1$ so that we can replace  $\sigma^z$ as the constant $\pm 1$.
Therefore,  the effective Hamiltonian becomes
\begin{eqnarray}
 \hat{H}_{eff} 
=\omega_{az} \hat{a}^\dagger \hat{a}
+\omega_{bz}  \hat{b}^\dagger \hat{b}
+g_{2,z}( \hat{a} \hat{b}^\dagger + \hat{a}^\dagger \hat{b}) +C_z \ \ , \nonumber\\
\omega_{az}\equiv \omega_a +g^2\gamma_a\sigma^z \ \ , \ \ 
\omega_{bz}\equiv \omega_b +g^2\gamma_b\sigma^z \ \ , \nonumber\\
 g_{2,z}  \equiv   \frac{1}{2}g^2( \gamma_a+\gamma_b) \sigma^z \ \ , 
 \nonumber\\
C_z \equiv  \frac{1}{2}\{\Omega+g^2( \gamma_a+\gamma_b)\} \sigma^z
+\frac{1}{2}g^2( \gamma_a+\gamma_b)
\ \ .\label{Ap4a}
\end{eqnarray}

Since $ \hat{a}$ couples with $\hat{b}^\dagger $,
 we make  $ \hat{H}_{eff} $ diagonal by
new operators $\alpha,\beta$.
First we calculate the products of $ \hat{a},\hat{b}$ by  $ \hat{\alpha}, \hat{\beta}$.
\begin{eqnarray}
\hat{a}= \cos\theta \hat{\alpha} +\sin\theta \hat{\beta}
\ \ , \ \
\hat{b}= -\sin\theta \hat{\alpha} +\cos\theta \hat{\beta}
\ \ , \nonumber\\
\hat{a}^\dagger \hat{a}
= \cos^2\theta  \hat{\alpha}^\dagger \hat{\alpha} +\sin^2\theta \hat{\beta}^\dagger \hat{\beta}+\nonumber\\
+\cos\theta \sin\theta( \hat{\alpha}^\dagger \hat{\beta}+\hat{\beta}^\dagger  \hat{\alpha})
\ \ , \nonumber\\
\hat{b}^\dagger \hat{b}
= \sin^2\theta  \hat{\alpha}^\dagger \hat{\alpha} +\cos^2\theta \hat{\beta}^\dagger \hat{\beta}-\nonumber\\
-\cos\theta \sin\theta( \hat{\alpha}^\dagger \hat{\beta}+\hat{\beta}^\dagger  \hat{\alpha})
\ \ , \nonumber\\
\hat{a}^\dagger \hat{b}
= \cos\theta\sin\theta (- \hat{\alpha}^\dagger \hat{\alpha} +\hat{\beta}^\dagger \hat{\beta})+\nonumber\\
+\cos^2\theta \hat{\alpha}^\dagger \hat{\beta}
 -\sin^2\theta \hat{\beta}^\dagger  \hat{\alpha}
\ \ .\label{Ap4b}
\end{eqnarray}
Then we calculate  $H_{eff} $ by $\hat{\alpha} , \hat{\beta}$.
\begin{eqnarray}
 \hat{H}_{eff} =\hskip 2cm \nonumber\\
=\omega_{az}\{ \cos^2\theta  \hat{\alpha}^\dagger \hat{\alpha} +\sin^2\theta \hat{\beta}^\dagger \hat{\beta}+\nonumber\\
+\cos\theta \sin\theta( \hat{\alpha}^\dagger \hat{\beta}+\hat{\beta}^\dagger  \hat{\alpha})\}+
\nonumber\\
+\omega_{bz} \{  \sin^2\theta  \hat{\alpha}^\dagger \hat{\alpha} +\cos^2\theta \hat{\beta}^\dagger \hat{\beta}-\nonumber\\
-\cos\theta \sin\theta( \hat{\alpha}^\dagger \hat{\beta}+\hat{\beta}^\dagger  \hat{\alpha})\}+
\nonumber\\
+g_{2,z}\{2\cos\theta\sin\theta (- \hat{\alpha}^\dagger \hat{\alpha} +\hat{\beta}^\dagger \hat{\beta}) +\nonumber\\
+(\cos^2\theta  -\sin^2\theta)(\hat{\alpha}^\dagger \hat{\beta}+
 \hat{\beta}^\dagger  \hat{\alpha})\}
 +C_z 
\ \ .\label{Ap4c}
\end{eqnarray}
In order to  eliminate the term of $\hat{\alpha}^\dagger \hat{\beta}$, we make
 $$\tan(2\theta)=\frac{2g_{2,z}}{   \omega_{bz}-\omega_{az}}     \ \ . $$
 Therefore, we obtain
 \begin{eqnarray}
 \hat{H}_{eff} 
=\omega_{az}\{ \cos^2\theta  \hat{\alpha}^\dagger \hat{\alpha} +\sin^2\theta \hat{\beta}^\dagger \hat{\beta}\}+
\nonumber\\
+\omega_{bz} \{  \sin^2\theta  \hat{\alpha}^\dagger \hat{\alpha} +\cos^2\theta \hat{\beta}^\dagger \hat{\beta}
\}+
\nonumber\\
+2g_{2,z}\{\cos\theta\sin\theta (- \hat{\alpha}^\dagger \hat{\alpha} +\hat{\beta}^\dagger \hat{\beta})\}
 +C_z =
 \nonumber\\
=\omega_{\alpha} \hat{\alpha}^\dagger \hat{\alpha} 
+\omega_{\beta}  \hat{\beta}^\dagger \hat{\beta}+C_z 
\ \ .\label{Ap4d}
\end{eqnarray}
Here 
\begin{eqnarray}
  \omega_\alpha=\frac{\omega_{az}+\omega_{bz}}{2}
- \frac{1 }{2}\sqrt{(\omega_{bz}-\omega_{az})^2+4g_{2,z}^2}
   \ \ , \nonumber\\
   \omega_\alpha=\frac{ \omega_{az}+\omega_{bz}}{2}
+ \frac{1 }{2}\sqrt{(\omega_{bz}-\omega_{az})^2+4g_{2,z}^2}  
\ \ .\label{Ap4e}
\end{eqnarray}
\vskip 0.5cm
\noindent
{ \bf C.3  Time Evolution of Number Operator}
\vskip 0.3cm
\noindent
We will examine the time evolution in the effective Hamiltonian (\ref{Ap4a}).
An operator $\hat{O}$ at time $t$ is defined by
$$ \hat{O}(t)\equiv e^{i \hat{H} t}  \hat{O}(0)  e^{-i \hat{H} t} \ \ . $$

We will  calculate $\langle   \Psi|\hat{a}^\dagger(t)\hat{a}(t) |\Psi\rangle $.
  \begin{eqnarray}
  \hat{a}(t)=\cos\theta \hat{\alpha}(t) +\sin\theta \hat{\beta}(t)=\nonumber\\
=\cos\theta \hat{\alpha}(0) e^{-i\omega_\alpha t}
+\sin\theta \hat{\beta}(0) e^{-i\omega_\beta t}  =\nonumber\\
=\cos\theta e^{-i\omega_\alpha t}\{ \cos\theta \hat{a}(0) -\sin\theta \hat{b}(0)\}
+\nonumber\\
+\sin\theta e^{-i\omega_\beta t} \{ \sin\theta \hat{a}(0) +\cos\theta \hat{b}(0)\}
 \ \ .\label{Ap6a}
\end{eqnarray}
 \begin{eqnarray}
\hat{a}(t)^\dagger\hat{a}(t)=
[\cos\theta e^{-i\omega_\alpha t}\{ \cos\theta \hat{a}(0) -\sin\theta \hat{b}(0)\}
+\nonumber\\
+\sin\theta e^{-i\omega_\beta t} \{ \sin\theta \hat{a}(0) +\cos\theta \hat{b}(0)\}]^\dagger\times
 \nonumber\\
\times [\cos\theta e^{-i\omega_\alpha t}\{ \cos\theta \hat{a}(0) -\sin\theta \hat{b}(0)\}
+\nonumber\\
+\sin\theta e^{-i\omega_\beta t} \{ \sin\theta \hat{a}(0) +\cos\theta \hat{b}(0)\}]=
\nonumber\\
=\cos^2\theta \{ \cos\theta \hat{a}(0) -\sin\theta \hat{b}(0)\}^\dagger\times
\nonumber\\
\times\{ \cos\theta \hat{a}(0) -\sin\theta \hat{b}(0)\}+\nonumber\\
+
\sin^2\theta \{ \sin\theta \hat{a}(0) +\cos\theta \hat{b}(0)\}^\dagger
\nonumber\\
 \times\{ \sin\theta \hat{a}(0) +\cos\theta \hat{b}(0)\} + \nonumber\\
+ [ \cos\theta \sin\theta\{ \cos\theta \hat{a}(0) -\sin\theta \hat{b}(0)\}^\dagger
\times \nonumber\\
\times \{ \sin\theta \hat{a}(0) +\cos\theta \hat{b}(0)\}
 e^{+i( \omega_\alpha-\omega_\beta)t}+c.c]
\ \ .\label{Ap6b}
\end{eqnarray}
For  initial state $|\Psi\rangle $,
we have the initial values for the number operators.
\begin{eqnarray}
\langle \Psi |   \hat{a}^\dagger(0) \hat{a}(0) |\Psi\rangle =n_a  \ \ , \nonumber\\
\langle \Psi |   \hat{b}^\dagger(0) \hat{b}(0) |\Psi\rangle =n_b \ \ , \nonumber\\
\langle \Psi |   \hat{a}^\dagger(0) \hat{b}(0) |\Psi\rangle =n_{ab}
\ \ .\label{Hf2a}
\end{eqnarray}
Therefore, we obtain
  \begin{eqnarray}
  \langle   \Psi |\hat{a}^\dagger(t)\hat{a}(t)|  |\Psi\rangle\hskip 2cm \nonumber\\
 =(\cos^4\theta+\sin^4\theta)n_a+2\cos^2\theta\sin^2\theta n_b
  +\nonumber\\
 +2(-\cos^2\theta+\sin^2\theta)\cos\theta\sin\theta (n_{ab} +n_{ab}^*)+
 \nonumber\\
+ \{2\cos^2\theta \sin^2\theta(n_a-n_b) + \hskip 1cm\nonumber\\
+\cos\theta \sin\theta (\cos^2\theta -\sin^2\theta )(n_{ab} +n_{ab}^*)  \}
\times \nonumber\\
 \times \cos\{( \omega_\alpha-\omega_\beta)t\}  +\hskip 2cm\nonumber\\
 +i\cos\theta \sin\theta (n_{ab} -n_{ab}^*)  
\sin\{( \omega_\alpha-\omega_\beta)t\}
\ \ .\label{Ap6d}
\end{eqnarray}

 %

For numerical discussions, we assume $n_{ab}=0$ for  simplicity.
In this case  the time evolution is given by
  \begin{eqnarray}
\langle \Psi |\hat{a}^\dagger(t)\hat{a}(t)  |\Psi\rangle 
 =(\cos^4\theta+\sin^4\theta)n_a+2\cos^2\theta\sin^2\theta n_b
 \nonumber\\
+ 2\cos^2\theta \sin^2\theta(n_a-n_b) 
 \cos\{( \omega_\alpha-\omega_\beta)t\} \ \ , \nonumber\\
   \langle   \Psi |\hat{b}^\dagger(t)\hat{b}(t)  |\Psi\rangle 
 =(\cos^4\theta+\sin^4\theta)n_b+2\cos^2\theta\sin^2\theta n_a
 \nonumber\\
+ 2\cos^2\theta \sin^2\theta(n_b-n_a) 
 \cos\{( \omega_\alpha-\omega_\beta)t\} 
\ \ .\label{Fg1a}
\end{eqnarray}
Note the following  conservation of the total
photon number.
$$\langle\Psi |\hat{a}^\dagger(t)\hat{a}(t)  |\Psi\rangle +\langle\Psi |\hat{b}^\dagger(t)\hat{b}(t)  |\Psi\rangle =n_a+n_b \ \ .$$

\end{document}